# Developing a Machine Learning-Based Clinical Decision Support Tool for Uterine Tumor Imaging

Darryl E. WRIGHT PhD
Rochester, Minnesota
Radiology, Mayo Clinic

Adriana V. GREGORY MS
Rochester, Minnesota
Radiology, Mayo Clinic

Deema ANAAM MBBS
Rochester, Minnesota
Radiology, Mayo Clinic

Sepideh YADOLLAHI MBBS
Rochester, Minnesota
Radiology, Mayo Clinic

Sumana RAMANATHAN MS
Rochester, Minnesota
Radiology, Mayo Clinic

Kafayat A. OYEMADE MD
Rochester, Minnesota
Obstetrics and Gynecology, Mayo Clinic

Reem ALSIBAI MBBS
Rochester, Minnesota
Obstetrics and Gynecology, Mayo Clinic

Heather HOLMES BS
Rochester, Minnesota
Radiology, Mayo Clinic

Harrison GOTTLICH BS
Rochester, Minnesota
Radiology, Mayo Clinic

Cherie-Akilah G. BROWNE BS
Rochester, Minnesota
Obstetrics and Gynecology, Mayo Clinic

Sarah L. COHEN RASSIER MD
Rochester, Minnesota
Obstetrics and Gynecology, Mayo Clinic

Isabel GREEN MD
Rochester, Minnesota
Obstetrics and Gynecology, Mayo Clinic

Elizabeth A. STEWART MD
Rochester, Minnesota
Obstetrics and Gynecology, Mayo Clinic

Hiroaki TAKAHASHI MD PhD
Rochester, Minnesota
Radiology, Mayo Clinic

Bohyun KIM MD
Rochester, Minnesota
Radiology, Mayo Clinic

Shannon LAUGHLIN-TOMMASO MD
Rochester, Minnesota
Obstetrics and Gynecology, Mayo Clinic

Timothy L. KLINE PhD
Rochester, Minnesota
Radiology, Mayo Clinic

The authors report no conflict of interest.

This project was partially supported by the Center for Individualized Medicine's Imaging Biomarker program at Mayo Clinic.

Timothy Lee Kline, PhD
Telephone: 507-284-2804
Fax: 507-284-8996
Email: Kline.Timothy@mayo.edu
Address: 200 First St SW, Rochester, MN 55905, USA

Word count: 4138

**Abstract**

Background: Uterine leiomyosarcoma (LMS) is a rare but aggressive malignancy. On imaging, it is difficult to differentiate LMS from, for example, degenerated leiomyoma (LM), a prevalent but benign condition. Failed diagnosis of LMS can have dire consequences. There is evidence that texture analysis of uterine tumors (UT) on MRI can assist in differentiation. As a first step texture analysis relies on the segmentation of presumed UTs.

Objective(s): We aimed to develop an automated approach to the segmentation of UTs and investigate the utility of radiomic features for distinguishing types.

Study Design: We curated a data set of 115 axial T2-weighted MRI images from 110 patients (mean [range] age=45 [17-81] years) with UTs that included five different tumor types. These data were randomly split stratifying on tumor volume into training (n=85) and test sets (n=30). An independent second reader (reader 2) provided manual segmentations for all test set images. To automate segmentation, we applied nnU-Net and explored the effect of training set size on performance by randomly generating subsets with 25, 45, 65 and 85 training set images. We evaluated the ability of radiomic features to distinguish between types of UT individually and when combined through feature selection and machine learning.

Results: Using the entire training set the mean [95% CI] fibroid DSC was measured as 0.87 [0.59-1.00] and the agreement between the two readers was 0.89 [0.77-1.0] on the test set. When classifying degenerated LM from LMS we achieve a test set F1-score of 0.80. Classifying UTs based on radiomic features we identify classifiers achieving F1-scores of 0.53 [0.45, 0.61] and 0.80 [0.80, 0.80] on the test set for the benign versus malignant, and degenerated LM versus LMS tasks.

Conclusion(s): We show that it is possible to develop an automated method for 3D segmentation of the uterus and UT that is close to human-level performance with fewer than 150 annotated images. For distinguishing UT types, while we train models that merit further investigation with additional data, reliable automatic differentiation of UTs remains a challenge.

## Introduction

Uterine leiomyoma (LM) is the most common gynecologic neoplasm with nearly 80% of women having at least one LM by age 50 [1]. The incidence of LM increases with age until menopause and is higher among African American women and those with a family history. By contrast, uterine leiomyosarcoma (LMS) is a rare tumor comprising 2-5% of all uterine cancer that often demonstrates aggressive behavior, local recurrence, and hematogenous and peritoneal dissemination [2-6].

When there are concerns of malignancy or as part of treatment planning, uterine magnetic resonance imaging (MRI) is often performed. The reported distinguishing MRI features of LMS include nodular border, presence of hemorrhage or necrosis, and diffusion restriction [7-9]. Degenerated LM, leiomyoma variants, and other uterine neoplasms make the distinction between LM and LMS on imaging even more difficult. Different types of spontaneous degeneration can occur, including hyaline, cystic, hemorrhagic and myxoid degeneration. Cellular LM, a histologic leiomyoma variant, shows increased cellularity relative to background myometrium. Uterine Smooth Muscle Tumors of Uncertain Malignant Potential (STUMP) show features that do not meet the criteria for LM or LMS, yet imaging diagnosis of STUMP tumors is difficult [10].

To overcome the considerable overlap in imaging findings of LM and LMS, we hypothesized that texture analysis with radiomics could provide additional quantitative metrics to improve tumor differentiation [11]. Radiomic analysis extracts image features including shape-based, intensity-based, and texture-based parameters from a region-of-interest. These extracted quantitative metrics have been applied across medical imaging modalities for detecting malignancy in applications such as lung nodules [12], pancreatic cysts [13], prostate tumors [14], and for the detection of Type-2 diabetes [15]. Radiomic features have also been shown to be predictive of underlying genetic mutations [16-18]. The first step for radiomic extraction is the delineation, or segmentation, of the region-of-interest from which radiomic features are to be extracted.

This study had two aims 1) to develop an automatic approach to delineate the tumor regions and 2) to investigate the potential of radiomic features to distinguish pathologically proven UT types.

## Materials and Methods

### *Dataset*

The Institutional Review Board approved this retrospective imaging study. Using institutional clinical data repository software (Advanced Cohort Explorer), pathologically proven LMS,

STUMP, cellular LM, and pathologically proven degenerating LM patients who had imaging within six months before surgery in any of the three sites of our institution since 1999 were included (n=294). To provide a comparison sample of non-degenerating LM, the institutional surgical database (SONX) was also searched for myomectomy with MRI findings of non-degenerating LM (n=32) within the prior 2 years. Finally, the institutional imaging database (MIDIA) was searched for MRI with leiomyoma between 2015-2020 (n=71). The search resulted in 397 cases. Excluding duplicates, cancers other than LMS, and those lacking prior research authorization 277 patients remained. Including only patients with MR imaging gave 164 cases. These images were manually reviewed to identify T2-weighted coronal series and ensure that pathology was available for each. For 47 cases no series matching these criteria was found or the T2-weighted axial series was deemed too low quality for inclusion. The final dataset included 115 images from 110 patients (Figure 1). These images were randomly split stratifying on UT type and volume into training (n=85) and test sets (n=30) while ensuring that images of the same patient were assigned to the same set. Table 1 shows the number of images and number of UTs by type.

Additional clinical information was gathered from Electronic Medical Records, including age, menstrual status and adenomyosis. We also curated details of the imaging parameters. Table 2 shows how these attributes are distributed across the training and test sets by UT type.

*Volumetric UT and uterus segmentation*

The UT and uterus were manually traced by trained readers (DA, HH) and reviewed/finalized by a radiologist (BJK, reader 1). A second radiologist (HT, reader 2) blinded to reader 1's work also provided manual segmentations for all test set images. All tracings were completed with the PKD-GUI (available online at https://github.com/TLKline/InstanceCystSeg) developed in our institution. The Graphical User Interface (GUI) allows each UT instance to be uniquely labeled and associated with its type. For each case, the final tracings were saved as two compressed NIFTI files, one containing the uterus mask and the other the UT masks. The UTs were classified as one of five classes –non-degenerating leiomyoma (NDLM), degenerating leiomyoma (DLM), cellular leiomyoma (CLM), smooth muscle tumors of uncertain malignant potential (STUMP), and leiomyosarcoma (LMS) – based on the pathology reports.

*Classification task definitions*

To test the hypothesis that radiomic features carry information relevant to distinguishing among UT types we focused on two clinically relevant binary classification tasks. The first combined the NDLM, DLM, and CLM into a "benign" class and the STUMP and LMS into a "malignant" class. The second aimed to differentiate DLM from LMS and was chosen since it is the most difficult to differentiate and commonly encountered clinically. In the supplementary material, we include results from three additional tasks CLM vs. LMS, CLM vs. NDLM, and STUMP vs. LMS. These were selected to target comparisons that are known to be difficult to distinguish on imaging or where there is little pre-existing evidence.

*Segmentation model development*

We used nnU-Net [19] to develop the automatic Artificial Intelligence (AI)-based segmentation model. nnU-Net is a deep learning segmentation method that is automatically self-configuring. The main advantage is that many aspects such as image preprocessing, network architecture, training strategy and post-processing that typically need to be tuned through systematic trial and error are instead based on fixed parameters, interdependent rules, and empirical decisions.

*Radiomic feature extraction*

Radiomic analyses were performed with the PyRadiomics library (version 3.0.1) [20]. Image voxel values were preprocessed by first clipping at the 99.9 percentile and then rescaling to the range 0 to 255 (Figure 2. panels A and B). A total of 107 radiomic features were extracted from the manual segmentation masks, which included 14 shape features, 18 first-order features, and 75 texture features. The texture features included 24 gray level co-occurrence matrix (GLCM), 16 gray level size zone matrix (GLSZM), 16 gray level run length matrix (GLRLM), 14 gray level dependence matrix (GLDM), and 5 neighboring gray-tone difference matrix (NGTDM) features. These features were only extracted from the region within the UT masks and each UT instance was treated individually.

*Machine learning*

All radiomic features were normalized using z-score normalization. Four machine learning algorithms were explored – logistic regression, support vector machine (SVM), random forests and extreme gradient boosting (XGBoost). In addition to training with the full feature set, the machine learning algorithms were combined with five feature selection methods – Minimum Redundancy Maximum Relevance (MRMR), top-K, Stability Selection, Least Absolute Shrinkage and Selection Operator (LASSO) and Principal Component Analysis (PCA). During training, a class weighting was applied to compensate for the imbalance in the number of training UTs depending on the classification task, for example, a weighting of 0.43:1.0 was applied for the DLM vs. LMS task. For an overview of machine learning best practices and applications in gynecologic imaging see Shrestha et al. (2022) [21].

*Statistical analyses*

The effect of training set size on the segmentation model performance was tested by randomly generating subsets with 25, 45, 65 and 85 training set images. Five-fold cross-validation on each subset was used to train nnU-Net. Model performance was measured in terms of the mean Dice similarity coefficient (DSC) measured across reader 1's test set segmentations. The DSC measures how closely the predicted segmentation matches that provided by reader 1, the higher the DSC the closer the match.

To estimate the value of additional training set images, the mean DSC on each subset was fitted with an exponential-plateau model. The point of plateau was defined as the smallest data set size that was within 1% of the maximum predicted DSC. An interobserver analysis was performed by comparing the mean DSC measured on the test set segmentations between the two readers and each reader compared to the automated segmentations.

To evaluate radiomic discriminative performance for individual features, first, for each feature a statistical significance test (Mann–Whitney *U* test) was performed across the training set for each of the five binary classification tasks. Second, each feature was treated as a classifier and a decision boundary was selected with three-fold cross-validation as the feature value that best separated the two classes ( ). For machine learning, three-fold cross-validation was applied to select the best combination of feature selection and learning algorithm along with their hyperparameters.

For both the individual feature classifiers and machine learning, classification performance was measured with the F1-score - chosen due to the imbalance in the number of examples for each class. Performance is reported as the mean [95% CI] F1-score. We refer to the mean F1-score across the training folds, the mean F1-score across the validation folds, and the mean F1-score of the three models trained for a particular feature selection method and learning algorithm combination on the test set as the training, validation, and test set performance respectively. The best approach according to the highest performance on the validation set was chosen for application to the test set.

As a benchmark, for each classification task, we report the performance of the naïve strategy of simply predicting all examples as members of the positive class. For the benign vs. malignant task this would be the F1-score resulting from classifying all images as malignant. A useful classifier should at least outperform this benchmark.

To place the results of the benign vs. malignant task in a clinical context we applied a Bayesian odds ratio analysis. We derived an estimate of the prior prevalence of malignancy among women with UT from the literature and used the false positive and true positive rates measured from a classifier’s performance on the test set. The result suggests how much our prior belief of malignancy should be updated given a prediction of malignancy by the classifier.

Figure 4 provides an overview of our methods and workflow with additional details given in the supplementary material.

## Results

### *Patient characteristics*

Of the 110 women whose images were included in the final dataset, the mean [range] age was 45 [17-81] years. Both age and menstrual status were significantly different between the benign (44 [17-81] years) and malignant patients (55 [25-78] years). This was anticipated given that LMS is more common in older, postmenopausal women.

### *Segmentation model*

Initial training of nnU-Net was done with the full training set (n=85), the test set DSC was measured as 0.87 [0.59, 1.00] for the uterus task and 0.92 [0.81, 1.00] for the combined uterus and UT task. The exponential-plateau model suggested a mean DSC of 0.90 for the uterus task could be achieved by increasing the training set size from 85 to 138 images (Figure 5 left). For the combined uterus and UT, the same analysis suggested that 61 images were sufficient to achieve the measured maximum mean DSC (Figure 5 right).

The interobserver analysis between the two radiology readers yielded test set DSC of 0.89 [0.77, 1.0] and 0.92 [0.81, 1.0] for the UT and combined uterus and UT tasks, respectively. Comparing nnU-Net to reader 2 gave 0.87 [0.59, 1.00] for the uterus and 0.91 [0.79, 1.00] for the combined uterus and UT, see Figure 6. Figure 7 shows a comparison of the nnU-Net segmentations to those of reader 1 for two examples drawn from the test set. The bottom panel of Figure 7 shows the worst-performing test set case.

### *Radiomic analysis*

For the benign versus malignant comparisons, 95 features were found to be significantly different including both texture and shape features (Table S3). For DLM versus LMS, 25 texture features were found to be significantly different but no shape features reached significance (Table S4). Tables S5-S7 show the additional classification tasks for other comparisons.

*Classification*

For DLM vs. LMS classification, the best-performing combination of machine learning algorithm and feature selection method was an SVM (C=1000; gamma=0.001; kernel=rbf) with top-K (K=25) feature selection with features ranked according to mutual information. The F1-score was 0.73 [0.47, 0.99] on the validation set and 0.80 [0.80, 0.80] on the test set. The naïve all LMS test set benchmark was 0.73. Treating individual features as classifiers, the median of the voxel signal intensities performed best with validation F1-score of 0.66 [0.42, 0.90] and 0.75 [0.68, 0.82] on the test set.

For the benign versus malignant classification task, the best performing combination of machine learning algorithm and feature selection method was an SVM (C=1000; gamma=10; kernel=linear) with LASSO feature selection using 10 features (see Figure S1). The F1-score was 0.56 [0.28, 0.83] on the validation set and 0.00 [0.00, 0.00] on the test set. The naïve all malignant test set benchmark was 0.065. Table 3 shows the 3 machine learning classifiers that achieved the highest performance on the validation set along with their corresponding training and test set performance. We found that the top two models seem to overfit the training set, however, the third best-performing model on the validation set (a Random Forest with 10 features selected with MRMR feature selection) measured similar F1-scores across the training, validation and test sets. When treating individual radiomic features as classifiers, the mean of voxel signal intensities performed best with validation F1-score 0.48 [0.15, 0.82] and 0.53 [0.45, 0.61] on the test set.

The best individual features as classifiers for each task are given in Table S8. Results for the remaining classification tasks are given in Table S9.

*Bayesian analysis*

We took the prior probability that a patient has LMS given that they have a UT as 0.002 (1 in 500) [8,22,23]. The prior prevalence of STUMP is less well understood, but we estimated it as half the prevalence of LMS which seems consistent with the relative proportions of STUMP to LMS in the literature [24,25] and our own data set. We, therefore, took the prior belief that a patient with a uterine tumor has a 0.003 or 1 in 333 chance the tumor is malignant. We applied Bayesian odds ratio analysis to the Random Forest classifier that achieved the third highest validation set F1-score on the benign vs. malignant task. This model incorrectly classified 11 benign UTs as malignant and correctly classified all malignant UTs, giving a false positive rate of 0.06 and a true positive rate of 1.00. The analysis suggests a prediction of malignancy from the model should update our belief a patient has a malignant UT to 0.046 or 1 in 22, a 10-fold difference. Given the uncertainty in the prior, we repeated the analysis assuming the prevalence of STUMP was either negligible (a prior of 0.002) or equal to that of LMS (a prior of 0.004) giving updated beliefs of malignancy of 1 in 32 or 1 in 17, respectively.

## Comment

### *Principal Findings*

In our study, we found that among 277 patients in the cohort, only 115 had usable magnetic resonance (MR) images. For some excluded patients, no axial MR sequence was available, or the quality was too poor for inclusion, however, for the majority only computed tomography (CT) imaging was available. With the available data, we were able to train a segmentation model that identified the uterus and uterine tumors (UTs) to a level competitive with that done manually by a radiologist. When measured against each reader's segmentations individually the model performed similarly to that reader and was close to the level of agreement between the two readers. Our analysis suggests that a segmentation model that is competitive with radiologists can be trained with fewer than 150 annotated images. The exponential plateau analysis suggested that the segmentation of UT could be improved by 3% by including an additional 53 examples in the training set. For the combined uterus and UT segmentation task, the same analysis suggested that with 85 training examples, we were already in the regime of diminishing returns and additional training examples would have little impact.

While most test set cases were similar to the radiologist readings, there was one significant outlier. This case had a smaller volume for both the uterus (72.90 ml) and uterine tumor (18.82 ml) and an imaging type which was underrepresented in the sample (a non-fat saturated image); both factors likely contributed to failure for this case highlighting the need for more standard imaging.

For classifying benign versus malignant UTs, we found that the approach selected by cross-validation failed on the test set and predicted all UTs as benign. The second-best performing model on the validation set similarly failed on the test set. Both models were linear classifiers combined with the Least Absolute Shrinkage and Selection Operator (LASSO) feature selection and appeared overfit, achieving perfect training set performance. In contrast, the third best performing classifier on the validation set performs similarly across the training, validation, and test sets. A data set containing more malignant examples could have helped prevent the selection of the top two classifiers suggested by cross-validation and selected instead the third top performer which appears to generalize better. We, therefore, favor the random forest model and expect its performance would be more robust on future data.

For the DLM vs. LMS task, we found that it was possible to outperform the naïve benchmark using either the median of voxel signal intensities or machine learning. Cross-validation favored selecting a Support Vector Machine-based classifier and we found this also provided the best test set performance.

*Results in the Context of What is Known*

The uterus segmentation performance in our study was similar to that of Shahedi et al. (2020) [26] who measured a Dice similarity coefficient (DSC) of 0.92 [0.83, 1.00] for the uterus on a test set of 20 MR images using a 3D U-Net model trained on 70 MR images to segment the uterus and placenta.

Using the random forest classifier for the benign versus malignant UTs task, we applied a Bayesian odds ratio analysis and found that a positive test should update our belief a patient has a malignant UT from 1 in ~333 to 1 in ~22. This analysis is based on 174 benign and only 6 malignant test set UTs. While these results are preliminary, we have identified a model against which future data can be tested to develop our confidence in its performance. These results also hint at the utility and motivation for developing such an approach, providing additional information that allows patients and surgeons to better estimate the risk of malignancy when weighing treatment options.

For the degenerated leiomyoma (DLM) vs. leiomyosarcoma (LMS) task, we found that key factors such as higher T2-weighted signal and heterogeneity were associated with LMS and important for classification [9,28-30], both of which can readily be identified by radiologists. Our classification performance for this task seems consistent with previous machine learning studies. In general, when following a systematic classification scheme radiologists outperform these automatic approaches [9].

*Clinical Implications*

LMS is a rare but aggressive malignancy that is difficult to differentiate from more common benign UTs. A reliable automated decision support tool for LMS would enable more comprehensive screening of the population. Although we have taken the first steps in this direction development of such a tool remains a challenge. We have identified models to apply to future data, however, this will be for validation purposes rather than clinical decision-making.

*Research Implications*

Our analysis suggests that future work should focus on improvements in UT segmentation whereas only marginal improvements are likely for uterus segmentation. Robust automated classification of UT types requires further work. In particular, the curation of additional cases would improve model development and confidence in model performance. For example, in the radiomic analysis, we found that the low number of malignant cases was a disadvantage in identifying a robust classifier. Additionally, schemes proposed to differentiate LMS from other UTs rely on several imaging modalities including T1-weighted, Diffusion Weighted and Apparent Diffusion Coefficient imaging [22,31]. Some of the curation burden for additional cases and images could be alleviated by our segmentation model. For new T2-weighted images, the segmentation model could be used to boost the number of examples available for the classification analysis. Models for segmentation of other modalities could be transfer-learned [32] from the T2-weighted segmentation model likely reducing the number of images requiring human segmentation from scratch while achieving similar performance. However, given how rare the malignant UTs are, a sufficient data set is likely beyond the capacity of any one institution and broader interinstitutional data sharing would help drive progress towards benefits for patients.

*Strengths and Limitations*

The strengths of this study are the curation of a high-quality data set and results for both segmentation and classification of UTs. We presented a highly performant segmentation model and reported the results of a comprehensive exploration of radiomic features. These results are set in the context of naïve benchmarks specific to the data set, but also benchmarks available in the literature where available.

However, the study also has a few limitations. It is retrospective over a long time period, and many patients came from outside institutions. This led to a lack of history such as previous treatment and the presence or absence of hereditary disease, but more importantly to great variation in imaging techniques. Given the rarity of the disease which limited the sample size, we chose to include as many images as possible despite the lack of standardization This low sample size limited the type and strength of analysis we could perform. Finally, schemes proposed to differentiate LMS from other UTs typically rely on several imaging modalities including T1-weighted, Diffusion Weighted and Apparent Diffusion Coefficient imaging [22,31]. The effort required to curate such a data set for the development of an automated approach precluded this approach and is exacerbated by the rare incidence of LMS and STUMP. These image series were also not available in all exams.

*Conclusions*

We have shown that it is possible to develop an automated method for 3D segmentation of the uterus and UTs in T2-weighted MR images that is close to human-level performance, even with fewer than 150 annotated images. We also conducted an extensive analysis of radiomic features extracted from UTs. While we trained models that merit further investigation with additional data, reliable automatic differentiation of UTs remains a challenge.

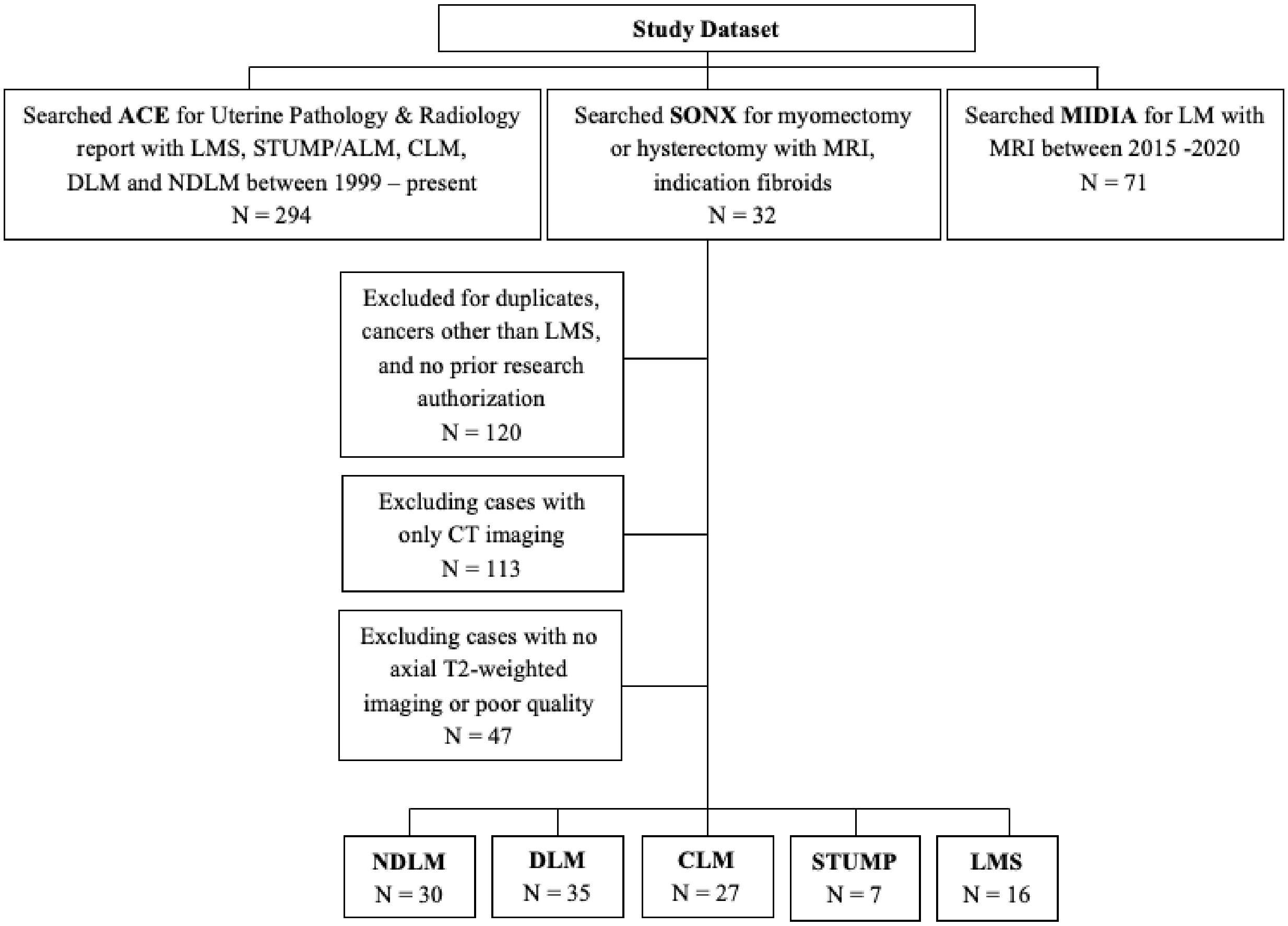

**Figure 1.** Schematic of data curation. N represents the number of images.

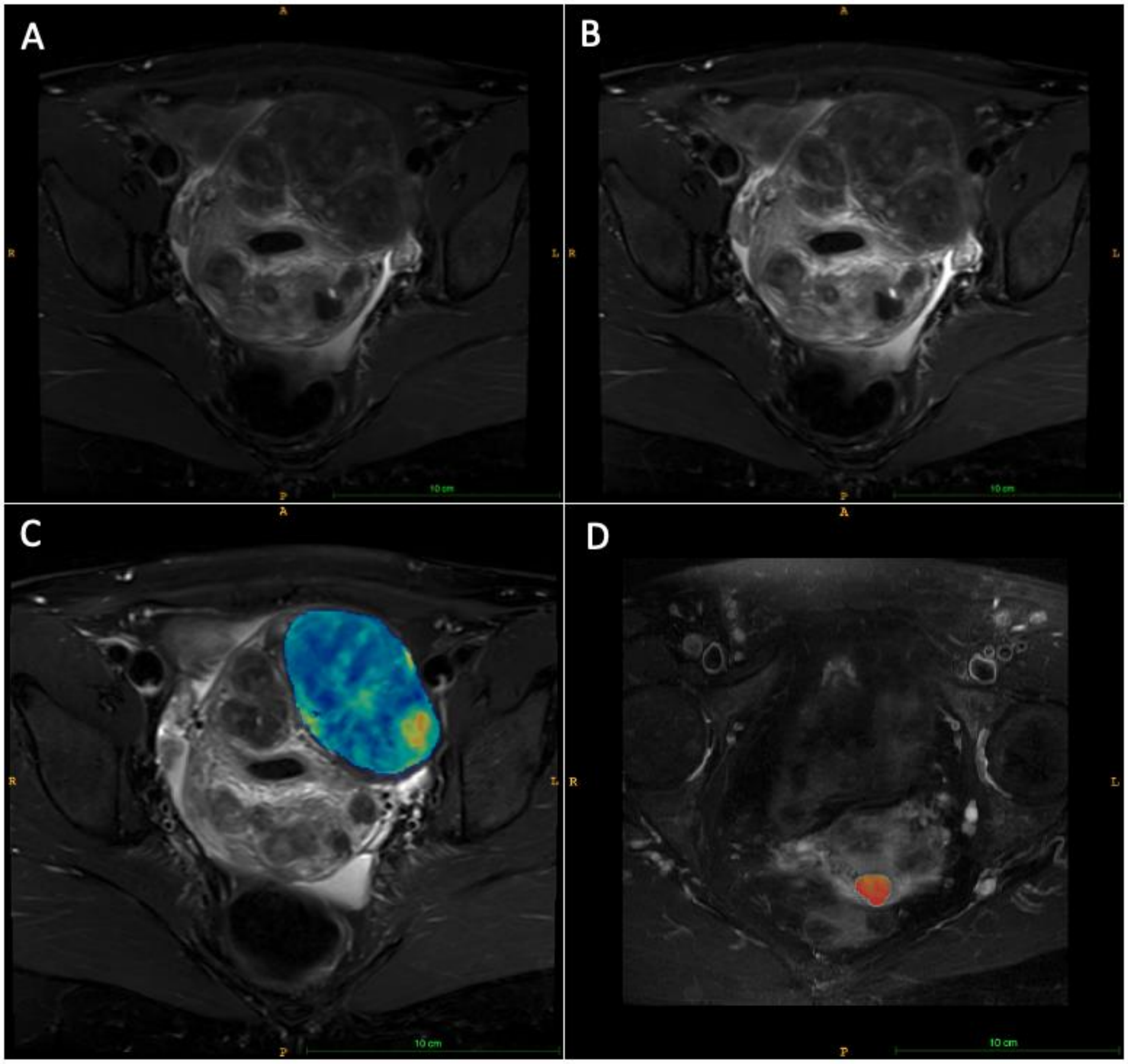

**Figure 2. A, B** show an example slice from an axial T2-weighted fat-saturated MR image before and after preprocessing, respectively. **C** The same as **B** but with the median of voxel signal intensities of a DLM overlaid. **D** Another preprocessed axial T2-weighted fat-saturated MR slice with the median of voxel signal intensities of an LMS overlaid.

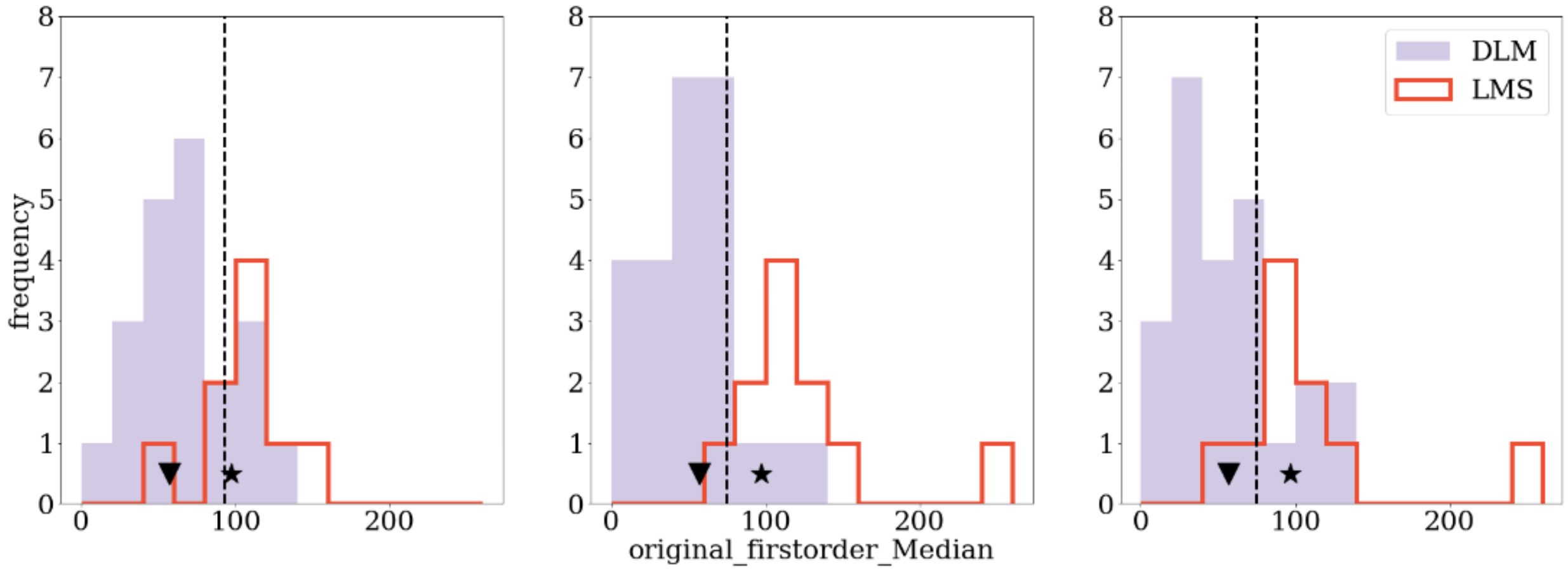

**Figure 3.** Individual radiomic features as classifiers. The histograms show the distribution of DLM and LMS training set UT according to the median of the voxel signal intensities in each of the three cross-validation folds. The decision boundaries (black vertical dashed lines) were chosen by selecting the threshold values that best separate the two classes in each fold. In all three folds, a case lying to the right of the decision boundary is classified as DLM. The star shows a test set case corresponding to the manual segmentation. The ground-truth label for this UT is DLM and is therefore incorrectly classified as LMS by all three decision boundaries. The majority vote is therefore also incorrect. The inverted triangle shows the same case but with the feature extracted from the AI-based segmentation shifting it to the other side of the decision boundary and correctly classifying it as DLM.

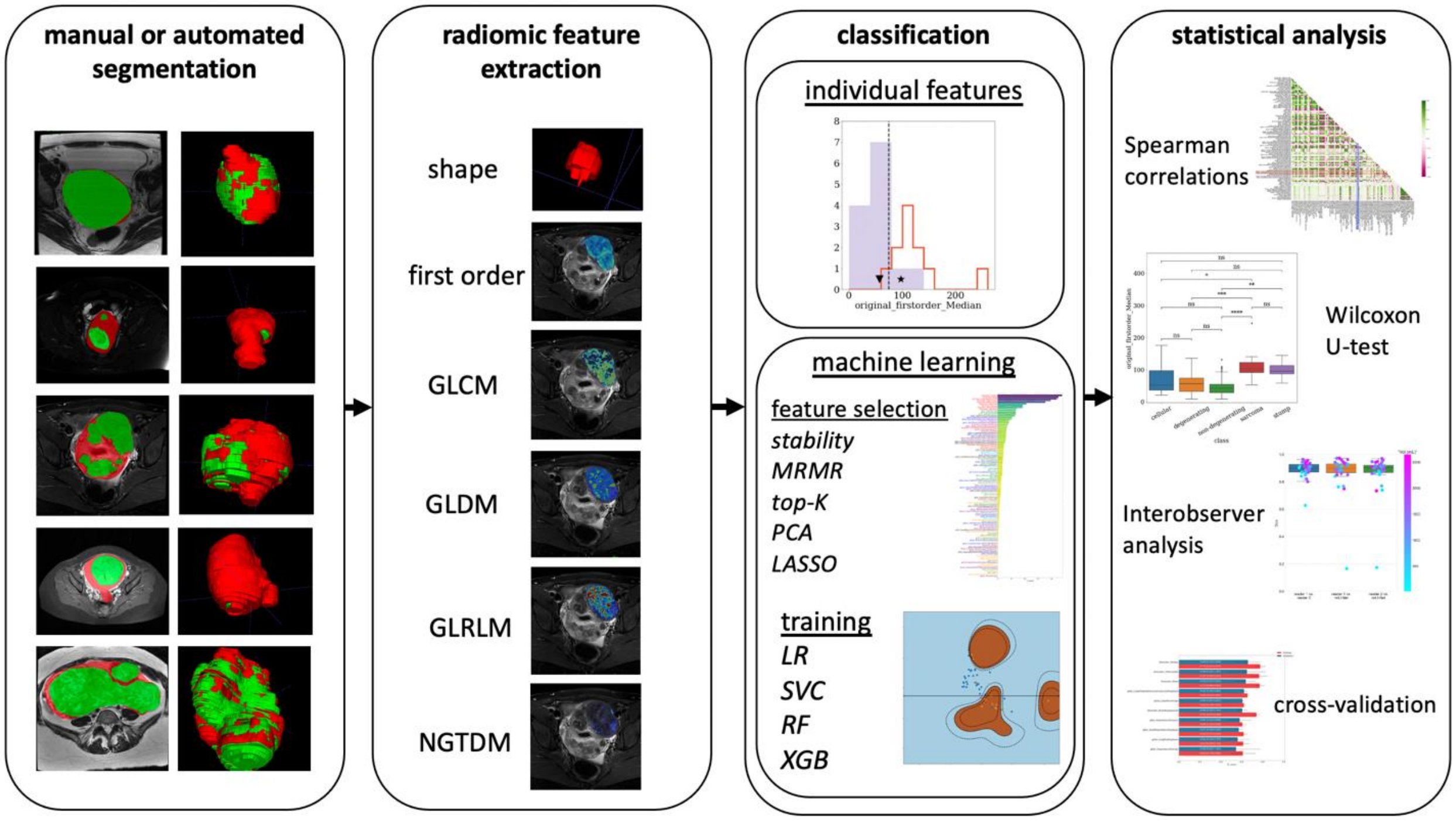

**Figure 4.** Methods and workflow. The uterus and UT are first segmented. From the segmented regions, we extract radiomic features. We explore the ability of individual radiomic features to distinguish UT types and use feature selection methods and machine learning to combine the most salient radiomic features to develop classifiers. We use statistical analyses to determine the best approaches and compare them to benchmarks. GLCM=Gray Level Co-occurrence Matrix, GLSZM=Gray Level Size Zone Matrix, GLRLM=Gray Level Run Length Matrix, GLDM=Gray Level Dependence Matrix, and NGTDM=Neighboring Gray-Tone Difference Matrix, MRMR=Maximum Relevance Minimum Redundancy, PCA=Principal Component Analysis, LASSO= Least Absolute Shrinkage and Selection Operator, LR=Logistic Regression, SVC=Support Vector Classifier, RF=Random Forest, XGB=eXtreme Gradient Boosting.

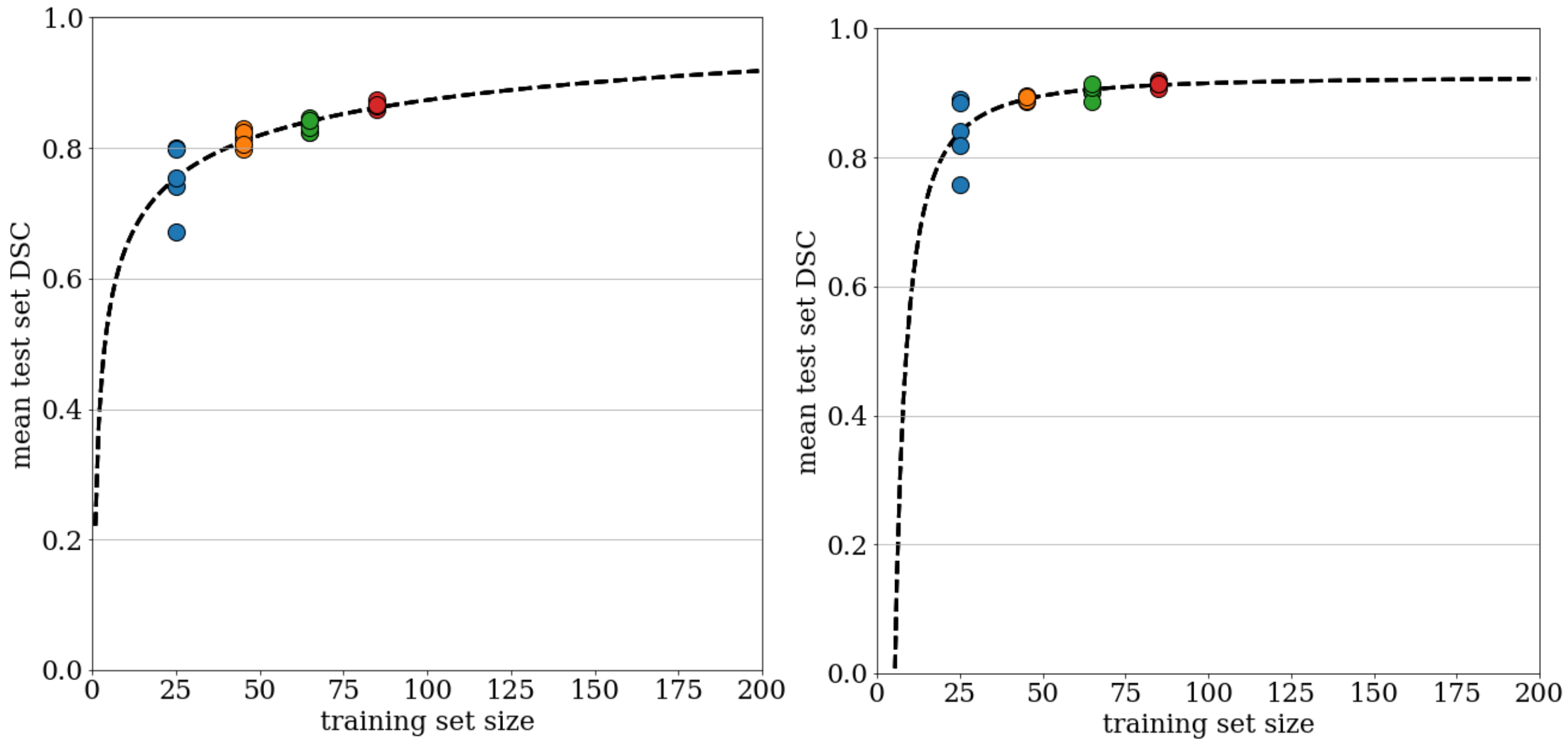

**Figure 5.** (left) The exponential-plateau analysis for the UT task. The model fit is shown as the dot-dashed blue line. The results suggest increasing the training set size to 138 images could boost performance to 0.90 DSC beyond which we expect performance to plateau. (right) The same but for the combined uterus and UT task. In this case, the performance has already plateaued, and additional training data will add no benefit.

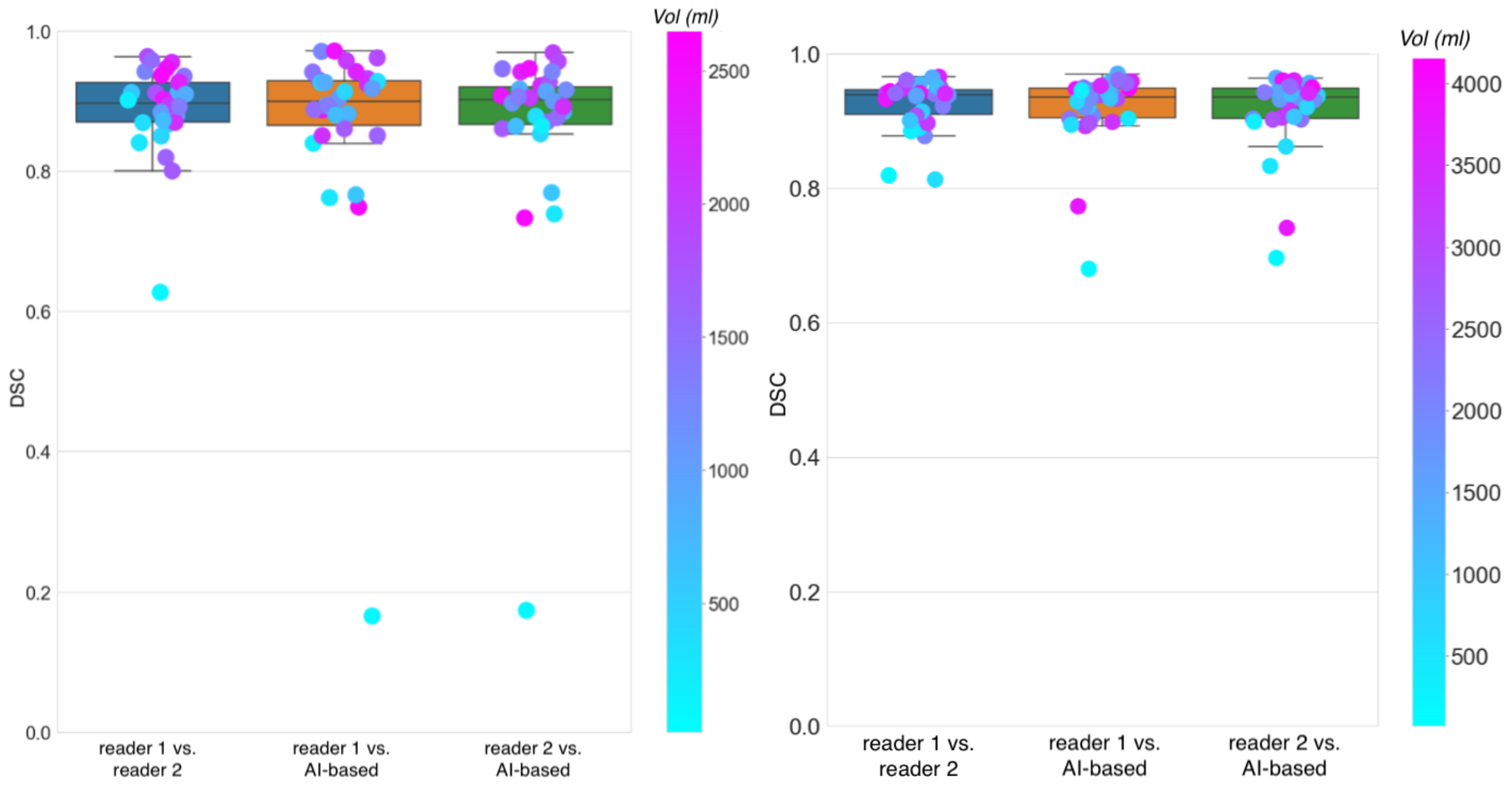

**Figure 6.** (left) Boxplot showing the interobserver analysis for the UT segmentation. The agreement between the two readers is compared to the agreement between the AI-based segmentations and each reader. (right) The same but for the combined uterus and UT task. For both tasks we find that the agreement between readers is similar to the agreement between the readers and the AI-based approach.

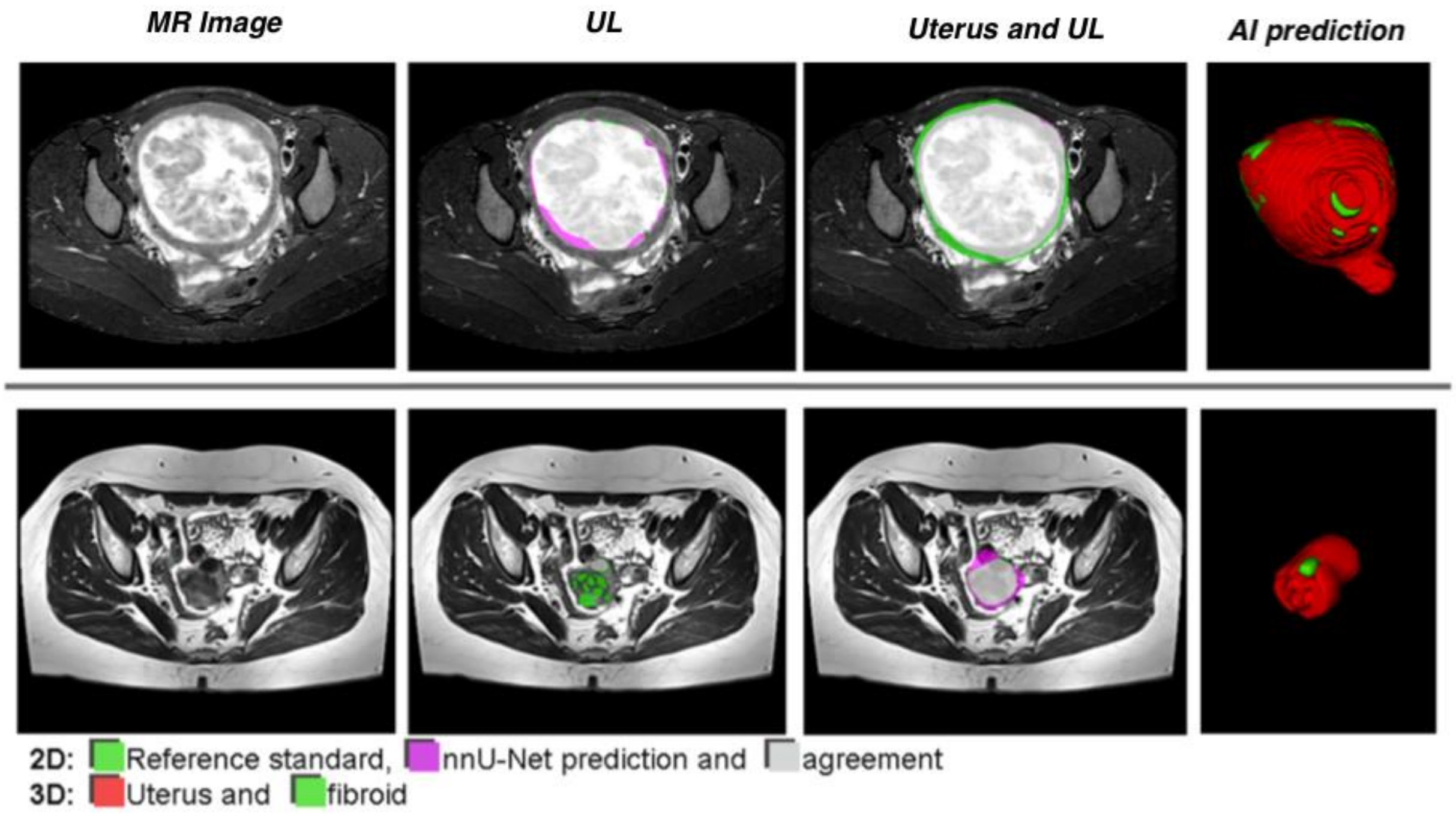

**Figure 7.** (top) An axial T2-weighted fat-saturated MR image of a 32 year-old female presenting a cellular leiomyoma. The DSC between reader 1 and the AI-based segmentation was 0.93 and 0.94 for the UT and uterus + UT, respectively. (bottom) An axial T2-weighted non-fat-saturated MR image of a 59 year-old female presenting multiple leiomyomas including a cellular leiomyoma. The DSC between reader 1 and the AI-Based segmentation was 0.17 and 0.68 for the UT and uterus + UT, respectively.

## Supplementary Material

*Dataset*

Age was included as the patient's age in integer years at the time of surgery. Menstrual status was encoded as a categorical variable with peri=0, pre=1, and post=2. Adenomyosis was similarly encoded as a categorical variable with yes=1, no=2, possible=3, probable=4, unknown=5. Whether or not an image was fat-saturated we encoded as 0=no and 1=yes.

*Segmentation model development*

The UL and uterus masks were combined into a single compressed NIfTI file with two classes. This was achieved by first combining all UL classes into a single UL mask and second, modifying the uterus mask to be the region of the uterus that does not contain UL (i.e. the difference between the new UL mask and the original uterus).

*Radiomic feature extraction*

For feature extraction, the preprocessed images were resampled to 0.75 × 0.75 × 5 mm (the mean in-plane × slice thickness calculated across the data set) with B-spline interpolation. Resampling of the masks relied on nearest-neighbor interpolation. A bin width of 25 was used.

We also extracted radiomic features using the test set segmentation masks predicted by nnU-Net. nnU-Net produces a single mask that includes all predicted fibroids. Individual UL were identified with a 3D connected components analysis and cross-matched with the manually segmented UL that resulted in the highest Dice similarity coefficient (DSC).

*Machine learning*

All radiomic features were normalized using z-score normalization across the training set. The variance and mean measured on the training set was applied to normalize the test set features.

Four machine learning algorithms were explored – logistic regression, support vector machine (SVM), random forests (RF) and XGBoost. In addition to the full feature set, these were combined with five feature selection methods – Minimum Redundancy Maximum Relevance (MRMR), top-K, Stability Selection, Least Absolute Shrinkage and Selection Operator (LASSO) and Principal Component Analysis (PCA). Tables S1 and S2 show the hyperparameters tried for each machine learning algorithm and feature selection method. For every feature selection method and its hyperparameter setting, we ran the full hyperparameter grid for each machine learning algorithm and evaluated performance with three-fold cross-validation. The same folds as in the individual feature analysis were used (see *Statistical analyses* below). During training a class weighting of 0.43:1.0 was applied to compensate for the DLM:LMS imbalance in the number of training fibroids. The best approach according to the highest mean F1-score was chosen for application to the test set.

*Individual feature classifiers*

Each feature was treated as a classifier and a decision boundary was selected with three-fold cross-validation as the feature value that best separated the two classes. Cross-validation was stratified to ensure that similar proportions of each class were assigned to each fold and that images of the same patient were included in the same fold. A classification was made by assigning a predicted class to a given MRI based on its position relative to the decision boundary (see Figure 3 for an illustrative example).

## Results

*Patient characteristics*

The mean [range] age of the patients was 45 [17-81] years. Both age and menstrual status were significantly different between the benign (44 [17-81] years) and malignant patients (55 [25-78] years) and found to be different for all other classification tasks except CLM versus NDLM. This is as anticipated given that LMS is more common in older, postmenopausal women. Age was found to be significantly different between the training and test set for the CLM versus NDLM task only, while menstrual status was not significant between the training and test sets for any task. Adenomyosis was also not significant for any task, however, for 40% of the patients adenomyosis was unknown.

*Imaging characteristics*

Magnetic field strength and manufacturer were found to be significantly different between classes for the benign versus malignant and CLM versus NDLM tasks. Whether or not an image was fat-saturated was found to be significant for the CLM versus LMS and CLM versus NDLM tasks.

*DLM vs LMS with nnU-Net segmentations*

Cross-matching the automatic and the manual instance segmentations resulted in median [range] DSC of 0.925 [0.002-0.973]. There were two cases where the cross-matching strategy failed with DSC of 0.014 and 0.002. In the first case the same automatic segmentation was cross matched with two different manually segmented UT. For the second, there were 17 NDLMs in addition to the three DLMs included in our analysis, making this case particularly complex to segment.

In the test set, no feature was found to be significantly different between the radiomics extracted from the manual segmentations and those extracted from the cross-matched nnU-Net UT segmentations. Applying the decision boundaries selected on the median voxel intensity to the radiomics extracted from the predicted nnU-Net segmentations gave F1-score of 0.78 [0.74, 0.83].

## Discussion

*Exponential plateau analysis*

It is important to note that this analysis assumes the same network architecture is trained and that future data will be drawn from the same distribution as our test set. If, for example, future data instead includes different imaging protocols then additional training data beyond that suggested by our analysis and including examples of that protocol would be beneficial.

*DLM vs LMS with nnU-Net segmentations*

Comparing the radiomics extracted from the manual and automated segmentations we found no statistically significant features suggesting that the quality of the automated segmentations is sufficient to expect robust radiomic extraction that would be similar to those extracted from manually generated equivalents. When the majority vote of the best performing ML classifier was applied to the automated segmentations, three labels were changed when compared to the manual approach. Two of these corresponded to the two cases where cross matching failed explaining the difference. This had the effect of flipping the incorrect LMS labels for these cases using the manual segmentations to the correct DLM labels when using the nnU-Net segmentations. This was serendipitous and not a strength of the automated segmentations over manual. The strategy for extracting instance segmentations from the automated approach could be improved in future by, for example, following the strategy of Gregory et al. (2021) [1]. The final case with a different label was correctly classified as a DLM with the manual segmentation, but incorrectly classified as LMS with the automated approach. Manual review showed that nnU-Net tended to over-segment the UT incorporating regions not included in the manual segmentations resulting in differences in extracted features that changed the classification.

*DLM vs. LMS*

Previously, Lakhman et al. (2017) [2] found that LMS was associated with higher signal intensity and a higher standard deviation and kurtosis of the signal intensity compared to Atypical LM (ALM). We found no significant difference comparing LMS to DLM for standard deviation in our data set. We also found that a lower kurtosis was associated with LMS than DLM. As these are first order features the contradictions could be due to differences in image parameters (they use only non-fat-saturated images) or preprocessing. Similar to our results Nakagawa et al. (2019) [3] found a higher mean T2-weighted signal was associated with LMS. Their classifier included kurtosis of the T2-weighted signal, but it is unclear whether higher kurtosis is associated with DLM or LMS. Lakhman et al. (2017) [1] also found that texture analysis suggests LMS is more heterogeneous than ALM. In our data set, of the 25 features significantly different between the two classes, 18 were related to texture. Those that indicate heterogeneity, such as GLDM small emphasis and GLRLM short run length low gray level emphasis, tended to be higher for LMS, in agreement with their findings. Comparing our machine learning approach to their results that use intensity image-based features and self-tuning spectral clustering we achieve higher sensitivity (1.0 versus 0.70) and specificity (0.78 versus 0.71). However, when ≥ 3 of four qualitative features (nodular borders, intra-lesional hemorrhage, “T2 dark” area(s) and central unenhanced area(s)) were manually identified by readers they measured higher sensitivity (95-100%) and specificity (95-100%) suggesting that machine learning approaches currently underperform in comparison to expert human-level performance. Nakagawa et al. (2019) [3] report a higher AUROC (Area Under the Receiver Operating Curve) for their XGBoost model compared to two radiologists, although it seems they were not following a systematic classification scheme.

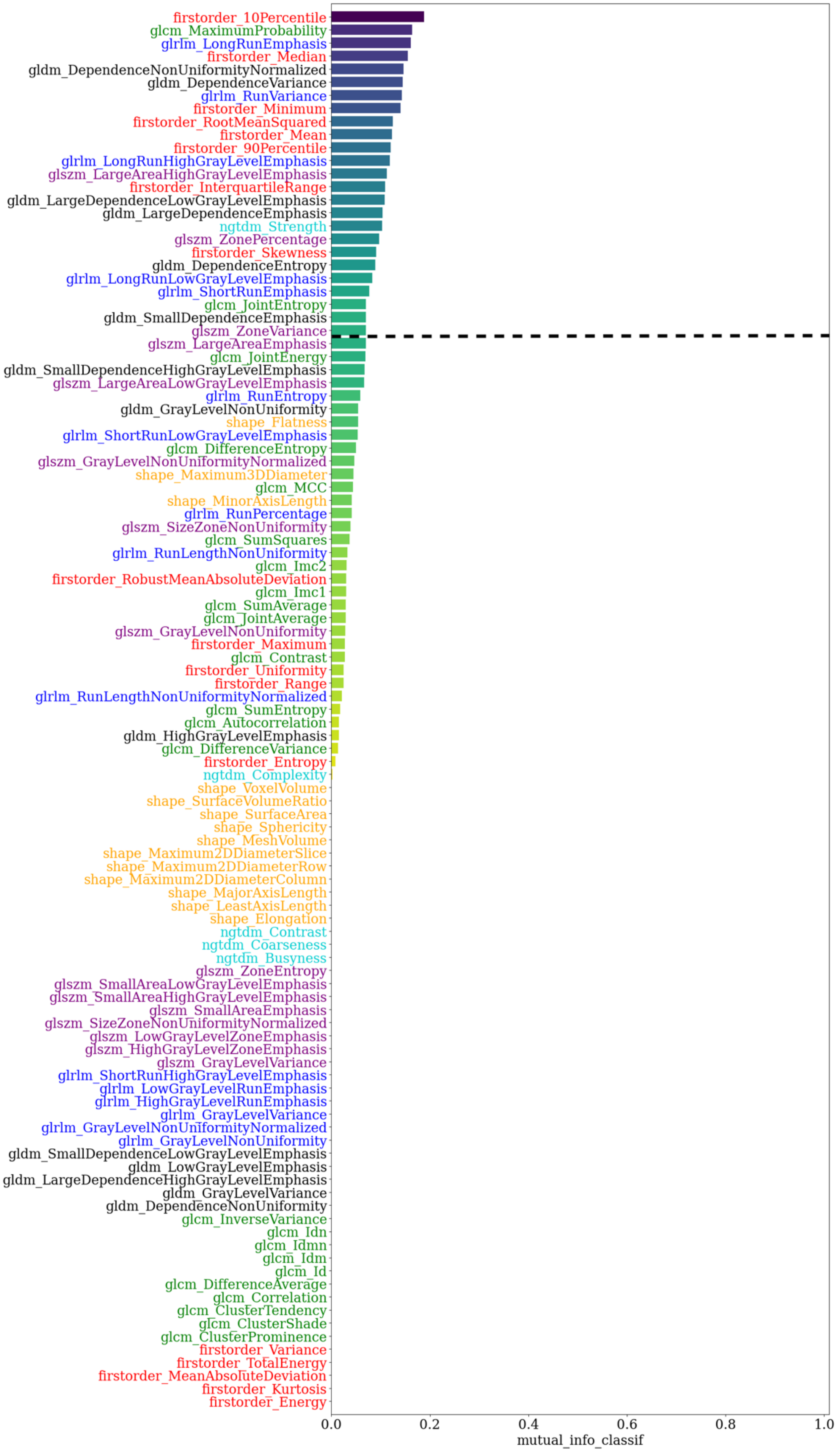

**Figure S1.** The 107 extracted radiomic features ranked according to mutual information. First order features are in red, Gray Level Cooccurrence Matrix (GLCM) features are in green, Gray Level Run Length Matrix (GLRLM) are in blue, Gray Level Size Zone Matrix (GLSZM) features are shown in purple, shape features are in orange, neighborhood Gray-Tone Difference Matrix (NGTDM) features are in cyan, Gray Level Dependence Matrix (GLDM) features are in black. The features above the horizontal dashed line are the 25 selected for training the best performing ML classifier.